\documentstyle[12pt]{article}
\begin{document}
\baselineskip 3.9ex
\def\be{\begin{equation}}
\def\ee{\end{equation}}
\def\ba{\begin{array}{l}}
\def\ea{\end{array}}
\def\bea{\begin{eqnarray}}
\def\eea{\end{eqnarray}}
\def\no#1{{\tt   hep-th/#1}}
\def\eq#1{(\ref{#1})}
\def\gap{\vspace{5ex}}
\def\lgap{\vspace{10ex}}
\def\sgap{\vspace{3ex}}
\def\del{\partial}
\def\o{{\cal O}}
\def\z{{\vec z}}
\def\W{{\cal W}}
\def\re#1{{\bf #1}}
\def\av#1{{\langle  #1 \rangle}}
\def\N{{\cal N}}
\def\S{{\cal S}}
\def\A{{\cal A}}
\def\g{{\gamma}}
\def\G{{\cal G}}
\def\ads{{AdS$_3$}}
\def\ni{\noindent}
\def\nn{\nonumber}

\renewcommand\arraystretch{1.5}

\begin{flushright}
TIFR/TH/99-23\\
IC/99/64\\
June 1999\\
hep-th/9906112
\end{flushright}
\begin{center}
\vspace{2 ex}
{\large{\bf Point Mass Geometries, Spectral Flow and
\ads-CFT$_2$ Correspondence}}\\
\vspace{3 ex}
Justin R. David$^{a,*}$, Gautam Mandal$^a$, 
Sachindeo Vaidya$^a$ and Spenta R. Wadia$^{a,b,\dagger}$ \\
$\,$\\
$^a${\sl Department of Theoretical Physics,
Tata Institute of Fundamental Research,}\\
{\sl Homi Bhabha Road, Mumbai 400 005, INDIA.} \\
$\,$ \\ 
$^b${\sl Abdus Salam International Centre for Theoretical
Physics,}
\\
{\sl Strada Costiera 11, Trieste 34014, ITALY.}

\vspace{10 ex}
\pretolerance=1000000
\bf ABSTRACT\\
\end{center}
\vspace{1 ex}

We discuss, in terms of the \ads-CFT$_2$ correspondence, a
one-parameter family of (asymptotically \ads) conical geometries which
are generated by point masses and interpolate between \ads\ and BTZ
spacetimes. We show that these correspond to spectral flow in $\N=
(4,4)$ SCFT$_2$ which interpolate between NS and R sectors.  Our
method involves representing the conical spaces as solutions of
three-dimensional supergravity based on the supergroup $SU(1,1|2)
\times SU(1,1|2)$. The boundary CFT we use is based on the D1/D5
system. The correspondence includes comparing the Euclidean free
energies between supergravity and SCFT for the family of conical
spaces including BTZ black holes.

\vfill
\sgap
\hrule
\vspace{0.5 ex}
\begin{flushleft}
\baselineskip 2ex
{\tenrm
e-mail:  justin,mandal,sachin,wadia@theory.tifr.res.in\\
$^*$ Address after August 31, 1999: 
Department of Physics, University of California,
Santa Barbara,\\ CA 93106, U.S.A. \\
$^\dagger$ Jawaharlal Nehru Center for Advanced
Scientific Research. Bangalore 560012, INDIA.}
\end{flushleft}
\baselineskip 3.9ex

\clearpage

\section{Introduction and Summary}

The recently conjectured AdS$_{d+1}$/CFT$_d$ correspondence has led to
a number of remarkable predictions for $\N=4$ Yang-Mills theory in
four dimensions. Considering the fact that such a correspondence
potentially {\em defines} quantum gravity in these backgrounds, one
should be able to describe various interesting dynamical phenomena in
gravity as well, including, {\em e.g.}  black hole formation.  The
case of $d=3$ is particularly attractive because both \ads\ and
CFT$_2$ are more tractable than their higher dimensional counterparts
and has nevertheless a rich physics content.

A particularly interesting class of solutions of three-dimensional
gravity with $\Lambda < 0$ are conical spacetimes \cite{DesJac} which
are generated by a point mass at the origin (see equation
(\ref{cmass})). If the mass exceeds a certain critical value, the
conical spacetime becomes a BTZ black hole. Indeed, it was observed by
\cite{PelSte} Peleg and Steif that in the context of a gravitational
collapse of a shell of variable rest mass, the resulting spacetime is
conical unless the rest mass exceeds a critical value. At this value
there is a critical phase transition and a BTZ black hole is formed.

It was pointed out by Brown and Henneaux \cite{BroHen} that the
conical spacetimes are ``asymptotically \ads'' and they belong to the
class of geometries whose ``asymptotic isometry group'' is {\em
Virasoro} $\times$ {\em Virasoro}. Therefore it seems reasonable to
expect, in the light of AdS/CFT correspondence, that these spaces
should correspond to boundary conformal field theories. It is already
known \cite{CouHen} that the ``end-points'' of the one-parameter
family of metrics ($\gamma=0,1$), namely the pure \ads\ and the
zero-mass BTZ black hole (see below equation \eq{btzmetric}),
correspond to the NS and R sectors of a $\N=(1,1)$ superconformal
field theory (SCFT).

If we go back to the origin of the \ads-CFT$_2$ correspondence
\cite{Maldacena}, namely to the near-horizon limit of the D1/D5
system, the natural supersymmetry of the SCFT is $\N=(4,4)$. (This
corresponds to a supergravity based on the group $\G= SU(1,1|2)\times
SU(1,1|2)$ \cite{deBoer}). In terms of such a SCFT, there is a natural
interpolation between the NS and R sectors called spectral flow
\cite{SchSei}. It was conjectured some time back \cite{Mandal} that
the conical spaces should correspond to spectral flow, the defect
angle being related to the parameter of the flow (see equation
\eq{etagamma}). One evidence was that the ADM mass in supergravity
matched exactly with $L_0 + \bar L_0$ in the SCFT (see equation
(\ref{landl})) for all values of the interpolating parameter. This
connection has also been mentioned recently in \cite{ChoNam}.

In incorporating conical spaces in the ambit of \ads-CFT$_2$
correspondence, the first step is to realize the conical spaces as
supersymmetric solutions of three-dimensional gravity, since the
boundary theory mentioned above is supersymmetric. If one tries to
generalize to conical spaces the work of \cite{CouHen} which embeds
BTZ and \ads\ in the framework of (1,1)-type\footnote{For the
definition of $(p,q)$-type \ads\ supergravity, please see
\cite{AchTow}. Note the difference with the notation $\N=(p,q)$ which
represents the number of supersymmetries in the left- and right-moving
sectors respectively of the SCFT.}  \ads, the conical spaces turn out
to be non-supersymmetric: the Killing spinors, in particular, are
quasiperiodic and hence not globally defined. If, however, one tries
to realize these as solutions of $SU(1,1|2) \times SU(1,1|2) $
supergravity, which is natural from the viewpoint of the boundary
theory, the Killing spinors become globally defined. This construction
also allows us to establish the main point of the bulk-to-boundary
correspondence in this case: since the same supergroup is represented
in the bulk theory and in the boundary theory, we can find the
operator in the supergroup which deforms the value of the spectral
parameter in the SCFT and the operator which changes the value of the
defect angle in the supergravity. We find that the same operator
causes the one-parameter flow in both cases, thus establishing the
correspondence between the spectral flow and conical spaces.

As additional evidence for this correspondence we compute the
Euclidean free energy of the family of spacetimes and compare with the
corresponding quantity in SCFT.  The BTZ free energy, which has
already been discussed in \cite{MalStr,Mano}, is shown to be
reproduced by SCFT based on the symmetric product $S_{Q_1Q_5}(T^4)$ at
high temperatures. The free energies of the conics are reproduced at
low temperatures.

As an off-shoot of our construction we find a vertex operator in the
SCFT which corresponds to creation of a point mass geometry in an
asymptotically \ads\ space. The scattering of such point masses
(including the cases when they form a black hole \cite{Steif}) is
naturally represented in terms of correlation function of such vertex
operators.

This paper is organized as follows. In section 2, we discuss the
conical spaces as solutions of three-dimensional supergravity (with
$\Lambda <0$) based on $\G=SU(1,1|2) \times SU(1,1|2)$.  In section 3
we make the identification with spectral flow in $\N=(4,4)$ SCFT. In
section 4 we compute the Euclidean free energies of these spaces from
supergravity.  In section 5 we compute the free energies from the SCFT
viewpoint and compare with the results of section 4. In the concluding
section (Sec. 6) we discuss the issue of scattering of the point
masses and black hole formation in terms of correlation functions in
the SCFT.

\section{Conical spaces as solutions of supergravity based on $\G$}

The action for three-dimensional supergravity (with cosmological
constant $\Lambda < 0$) based on $\G \equiv SU(1,1|2) \times
SU(1,1|2)$ is as follows \cite{Tan,Justin}: 
\bea
\label{action}
S &=& \frac{1}{16 \pi G^{(3)}_N} \int d^3 x
\left[\right. eR+\frac{2}{l^2}e \\ \nonumber 
-\epsilon^{\mu\nu\rho} \bar{\psi}_{\mu} {\cal D}_{ \nu}
\psi_{\rho} \!\!&-&\!\! 8l\epsilon^{\mu\nu\rho} (
A_{\mu}^i
\del_{\nu} A_{\rho}^i - \frac{ 4i\epsilon_{ijk} }{3}
A_{\mu}^i
A_{\nu}^j A_{\rho}^k ) \\  \nonumber
-\epsilon^{\mu\nu\rho} \bar{\psi'}_{\mu} {\cal D'}_{ \nu}
\psi'_{\rho} \!\!&+&\!\! 8l\epsilon^{\mu\nu\rho}(
A'^i_{\mu}
\del_{\nu} A'^i_{\rho} - \frac{ 4i\epsilon_{ijk} }{3}
A'^i_{\mu}
A'^j_{\nu} A'^k_{\rho})  \left. \right] \\  \nonumber
\eea
where $ l^2 = -1/\Lambda$, $G^{(3)}_N$ is the three-dimensional
Newton's constant, and ${\cal D}_{\nu } = \del_{\nu} +
\omega_{ab \nu }\gamma^{ab}/{4} - e_{a
\nu}\gamma^{a}/{(2l)} -
2 A_{ \nu}^i\sigma^i$ and ${\cal D'}_{\nu } =
\del_{\nu} + \omega_{ab \nu }\gamma^{ab} /{4} + e_{a
\nu}\gamma^{a}/{(2l)} -2 A'^i_{ \nu}\sigma^i$.  The basic fields
appearing in the lagrangian are the vierbein $e^a_\mu, \psi_\rho,
A^i_\mu, \psi'_\mu$ and $A'^i_\mu$.

The same three-dimensional supergravity can be obtained from type IIB
string theory compactified on $K=T^4 \times S^3$ (with constant flux
on $S^3$). Recall that the near-horizon geometry of the D1/D5 system
involves $K \times$\ads$\,$ whose (super)isometries are $\G$. Similarly
the near-horizon geometry of the five-dimensional black hole is $K
\times $ BTZ \cite{MalStr}.  This implies in an obvious fashion that
\ads\ and BTZ are solutions of \eq{action}. Furthermore, the
three-dimensional Newton's constant $G^{(3)}_N$ and the AdS radius $l$
are given by the following string theoretic expressions:
\bea
\label{actionparameters}
G^{(3)}_N &=& \frac{4 \pi^4 g^2_s}{V_4 l^3} \\
\nn
l^4 &=& \frac{16 \pi^4 g^2_s Q_1 Q_5}{V_4}
\\
\nn
\eea 
where $V_4$ is the volume of $T^4$ and $g_s$ is the string coupling
(we are working in the units $\alpha'=1$).

\vfill\eject

\ni\underbar{Conical spaces}

\sgap

As mentioned in the introduction, there is a one-parameter family of
classical solutions of three-dimensional gravity (with $\Lambda<0$)
discovered by Deser and Jackiw which are asymptotically \ads\ and
interpolate between the \ads\ and zero-mass BTZ solutions. These
represent geometries around a point mass $m, 0<m<1/(4G^{(3)}_N)$ 
which
create a conical singularity at the origin. The metric is given by
\be
\label{conicmetric}
ds^2= -dt^2 (\gamma + \frac{r^2}{l^2}) + dr^2 (\gamma + 
\frac{r^2}{l^2})^{-1} + r^2 d\phi^2,
\ee
It is easy to see that there is a conical singularity at the origin
$r=0$ with defect angle
\be
\label{defectangle}
\Delta\phi = 2\pi (1 - \sqrt{\gamma})
\ee
The parameter $\gamma$ varies from 0 to 1, and is related to the point
mass $m$ at the origin:
\be
\label{djmass}
m = \frac{1- \sqrt\g}{4 G_N^{(3)}} \label{cmass}
\ee
We will denote these spaces by $X_\g$. Note that the mass $m$ is the
DeserJackiw mass which is different from the BTZ definition of mass,
denoted by $M$, which is given by
\be
\label{btzmass}
M= -\g/(8 G_N^{(3)}) 
\ee

\sgap

\ni\underbar{BTZ and \ads\ as end-points:}

\sgap

Recall that the metric of pure \ads\ is given by
\be
\label{adsmetric}
ds^2= -dt^2 (1 + \frac{r^2}{l^2}) + dr^2 (1 + 
\frac{r^2}{l^2})^{-1} + r^2 d\phi^2
\ee
while the metric for BTZ black holes is given by
\be
\label{btzmetric}
ds^2 = -\left[ \frac{r^2}{l^2} - M + (\frac{J}{2r})^2 \right] dt^2
+ \left[ \frac{r^2}{l^2} - M + (\frac{J}{2r})^2 \right]^{-1} dr^2
+ r^2(\frac{-J}{2r^2} dt +d\phi )^2 
\ee
where $M$ and $J$ refer to the mass and the angular momentum of the BTZ
black hole ($J \le M$).

It is clear that the space $X_\g$ becomes \ads\ for $\g=1$, whereas
for $\g=0$ it becomes BTZ with $M=0$ ($M=0 \Rightarrow J=0$).

\sgap

We have already remarked that \ads\ and BTZ are supersymmetric
solutions of \eq{action}. We will now show that the entire family of
spaces $X_\g$ can be obtained as supersymmetric solutions of 
\eq{action}.

It is known \cite{IzqTow}\footnote{We thank P.Townsend for pointing
out reference \cite{IzqTow} to us.} how to embed the conical
spaces \eq{conicmetric} as supersymmetric solutions in $(2,0)$-type
\ads\ supergravity. The solution looks like:
\be
\label{wilsonlineabelian}
A_{\mu} dx^\mu =  -\frac{l}{2}\gamma d\phi
\ee
where $A_\mu$ is the $U(1)$ gauge field appearing in the graviton
supermultiplet.

The embedding into $\N=(4,4)$ is a straightforward extension of the
above:
\be 
\label{wilsonline}
A_{\mu}^3 dx^\mu = -\frac{l}{2}\gamma d\phi, \quad
A_{\mu}^\pm dx^\mu =0 
\ee 
where the superscripts $3,\pm$ refer to the R-parity
group $SU(2)$. 

It can be explicitly checked that the equations of motion following
from \eq{action}, as well as the Killing spinor equations ensuring
supersymmetry are satisfied by the above solution.  The equations of
motion reduce to those of the $U(1)$ problem \cite{IzqTow} with the
ansatz \eq{wilsonline}.  As far as the Killing spinors are concerned,
the solution in our case is a doublet constructed out of the solution
of the $U(1)$ problem. The Killing spinor equation in the present case
is 
\be
\label{killingspinorequation}
{\cal D}_{\nu } \epsilon  = \left[ \del_{\nu} +
\frac{\omega_{ab
\nu }\gamma^{ab}}{4} - \frac{1}{2l} e_{a \nu}\gamma^{a} +
\frac{i}{l}
A^3_{ \nu} \right] \epsilon =0
\ee
The solution is given by
\be 
\label{killingspinor}
\epsilon= \left( \begin{array}{c}
                 \epsilon_1 \\
                 \epsilon^*_1
              \end{array} \right),  
\ee
where $\epsilon^*_1$ is the complex conjugate of $\epsilon_1$, the
latter being the Killing spinor in the U(1) problem \cite{IzqTow}.

\sgap

\ni\underbar{Connection with $(1,1)$-type \ads\ supergravity:}

\sgap

We emphasize that extended supersymmetry is quite essential to
construct the conical spaces as supersymmetric solutions.  The Killing
spinors have holonomies under both the spin connection and the gauge
connection. Under either one of them the spinors are quasiperiodic,
corresponding to the fact that the $(1,1)$-type Killing spinors
constructed using the formulae of \cite{CouHen}, which see only the
spin connection, are quasiperiodic. With extended supersymmetry ({\em
i.e.}, $(p,q)$-type \ads\ supergravity, with either $p$ or $q$ or both
greater than 1) there is a $U(1)$ gauge field. The holonomy under
this gauge connection cancels that under the spin connection, making
the Killing spinors periodic and hence globally defined. The NS or R
boundary conditions of the spinors of \cite{CouHen} for $\g=1,0$ now
refer to the condition that the holonomy under the gauge connection,
represented by the Wilson line
\be
\label{defwilsonline}
\W= {\rm Tr} \exp[\frac{i}{l} \int A_{\mu} dx^\mu],
\ee
is $-{\bf 1}$ or ${\bf 1}$ respectively.

\sgap

\ni\underbar{Conics as one-parameter family of gauge transforms}

\sgap

It is clear from the above discussion that the family of solutions can
be parameterized uniquely by the value of the Wilson line $\W$
(the metric is fixed once we specify this). We can, therefore, change
from the classical solution $X_{\g=0}$ to $X_{\g}$ by making a ``gauge
transformation''
\be
\label{gaugetransformation}
A_\phi \to U^{-1}A_\phi U + i U^{-1}\del_\phi U, \quad
U(t,r,\phi) = \exp[i \zeta(\phi) T^3], \quad 
\zeta(\phi) =\phi\sqrt\g
\ee
The reason why the Wilson line $\W$ changes is that $U$ is not
single-valued. In other words, \eq{gaugetransformation}\ describes an
improper gauge transformation.

The equations of motion demand that the metric changes appropriately
from its value at $\g=0$ to the value at $\g$. It is interesting to
note that this change in the metric can also be understood as an
(improper) $SL(2,R)\times SL(2,R)$ gauge transformation. Following
\cite{Wittencs,AchTow} we combine the dribein $e^M_\mu$ and the spin
connection $\omega^M_{N\mu}$ into an $SL(2,R)\times SL(2,R)$ vector
potential $\A_\pm$. For the metric \eq{conicmetric}, 
it turns out to be
\be
\label{sl2rvalue}
\A_\pm \equiv \A^a_{\mu\pm}T^a dx^\mu =\frac{1}{2}
\left( \begin{array}{cc} 
                \pm d\rho &  -\sqrt\g e^{\mp \rho} dt_\pm  \\
                -\sqrt\g e^{\pm \rho}dt_\pm    & \mp d\rho \\
        \end{array} \right)
\ee
where $ r= \g \sinh \rho$ and $t_\pm = t \pm \phi$. One can show that
this can be transformed to the value at $\g=0$ by the following
$SL(2,R)\times SL(2,R)$ gauge transformation
\bea
\A_\pm  &\to& V_\pm^{-1} \A_\pm V_\pm + V_\pm^{-1} d V_\pm
\\ \nn
V_\pm &=& V_{1\pm} V_{2\pm}
\\ \nn
V_{1\pm} &\equiv& \left( \begin{array}{cc}
                                  e^{\mp \rho/2} & 0 \\
                                      0          & e^{\pm \rho/2}
                         \end{array} \right)   
\\ \nn
V_{2\pm} &\equiv& \exp[(1/2) \zeta(\phi) \sigma_1] 
\eea
$\zeta(\phi)$ being the same function as in \eq{gaugetransformation}.
Here $\sigma_1$ is the Pauli matrix. 

As emphasized before, the gauge transformation of the metric can be
regarded as a consequence of the gauge transformation
\eq{gaugetransformation}\ and the equations of motion. For comparison
with the boundary CFT, we note that the gauge transformation
\eq{gaugetransformation}\ is implemented in the quantum theory by the
operator
\be
\label{generator}
\widehat U  = \exp[i \sqrt\g \phi \widehat{J_3}]
\ee 
where $\widehat{J_a}, a=1,2,3$ are generators of the $SU(2)$ part of
$SU(1,1|2)$.

\section{Correspondence with spectral flow}

We have already remarked that the Brown-Henneaux Virasoro algabra can
be supersymmetrized \cite{CouHen} by embedding the \ads\ or BTZ
solutions in $\N=1$ supergravity, and that the realizations of the
superconformal algebra corresponding to the \ads\ and BTZ solutions
respectively map to the NS and R sector of the boundary conformal
field theory. In the context of extended supersymmetry, as we have
remarked above, the \ads\ and BTZ solutions correspond to a Wilson
line $\W$ equal to $-{\bf 1}$ or 
${\bf 1}$. In order to facilitate comparison, let us
define the gauge-invariantized fermion
\be
\label{wilsonlineu}
\tilde \psi \equiv U[A] \psi, \quad U[A]=\exp[\frac{i}{l} 
\int_{P_0}^P A] 
\ee 
The \ads\ and BTZ solution correspond to antiperiodic or periodic
$\tilde \psi$ respectively. It is this $\tilde \psi$ of supergravity
that maps to the fermion $\psi$ of CFT.

Now, we know that there is a one-parameter flow, called spectral flow,
between periodic and antiperiodic boundary conditions in the case of
$\N=4$ superconformal theory \cite{SchSei}. This corresponds to
quasiperiodic boundary condition on $\psi$:
\be
\psi(ze^{2\pi i}) = \exp(i\pi\eta) \psi(z)
\ee
Let us compare this flow with the one-parameter flow caused by the
Wilson line (Eqs. \eq{gaugetransformation} and \eq{generator}). We
note that the gauge-invariantized fermion $\tilde \psi$ satisfies,
under $\phi\to \phi + 2\pi$
\be
\tilde \psi(ze^{2\pi i},r\to \infty) = \exp(i \pi \sqrt\g)
\tilde \psi(z)
\ee
where $z= e^{i(t+ \phi)}$.  This suggests that the Hilbert space
corresponding to spectral flow is the realization of boundary SCFT for
the conical spaces, with the identification
\be
\label{etagamma}
\eta=\sqrt\g
\ee
The fact that this is the right correspondence follows by noting that
spectral flow in the SCFT is defined in terms of the generator
\cite{SchSei,Yu}
\be
\label{generatorCFT}
\widehat U  = \exp[i \sqrt\g \phi \widehat{J_3}]
\ee 
where $\widehat{J_3}$ is the $SU(2)$ R-parity current.  
Thus the $SU(2)$ generator for spectral flow \eq{generatorCFT}
is the same as the $SU(2)$ generator \eq{generator} in the bulk
that changes the conical defect angle. This is exactly as
it should be for the proposed AdS/CFT correspondence
to work.

The equality
of \eq{generator} and \eq{generatorCFT} under the AdS/CFT
correspondence proves our assertion.

A rather important consequence of the above correspondence is that we
know exactly what state in the CFT corresponds to a point mass in the
bulk that creates the conical singularity, namely it is the state
\be 
|\g \rangle \equiv \widehat U |0 \rangle,
\ee 
This suggests a boundary representation in terms of CFT vertex
operators of a point mass in the bulk and consequently a
representation of their scattering in terms of CFT correlation
in the sense of \cite{WittenAds,GubKlePol}. We
will make some more remarks on this in the concluding section.
Related remarks also appear in \cite{Martinec}\ in a somewhat
different context. 

We note that we are parameterizing the flow in supergravity by the
Wilson line only, by adopting the attitude that the metric gets fixed
as a consequence of the equation of motion.  For example, the ADM mass
for the metric corresponding to the Wilson line (\ref{wilsonline}) is
\be
\label{adm_mass}
M = - \g/8 G^{(3)}_N 
\ee
The counterart of this statement in CFT
is  that by performing a spectral flow
along $J^3$ we automatically change the value of $L_0$ 
on the ground state from 0 to
\be
\label{g_energy}
L_0 = -\frac{c}{24}\eta^2 
\ee
Similar remarks apply to $\bar L_0$. Using (\ref{actionparameters}) 
and the fact that $c=6Q_1 Q_5$ (see Section 5) for
the \ads\ and BTZ points, it is easy to see that 
\be
M= \frac{L_0 + \bar{L}_0}{l}. \label{landl}
\ee
We see that (\ref{adm_mass}) and (\ref{g_energy}) agree with
(\ref{landl}) if we use $\gamma = \eta^2$, which is the same condition
as in Eqn.  (\ref{etagamma}). This provides additional support to our
proposed correspondence. This matching has recently also been
mentioned in \cite{ChoNam}.

We will continue to explore this correspondence in the rest of the
paper in the context of the Euclidean free energy.

\section{The Euclidean free energy of asymptotically
\ads\ solutions}

In this section we will compute the Euclidean free energy of conical
spaces (also of \ads\ and BTZ) following the method of Gibbons and
Hawking \cite{GibHaw}. Let us recall that the free energy is given by
\be
\label{free}
Z \equiv \exp[-\beta F(\alpha)] = \int 
{\cal D}_\alpha[\hbox{fields}] \exp[-S]
\ee
where $S$ is now the Euclidean version of the action written in
\eq{action}.

\sgap

\ni\underbar{Boundary Conditions:}

\sgap

In the above equation, $\alpha$ denotes boundary conditions on the
fields. As noted in \cite{MalStr}, the boundary (corresponding to
$r\to \infty$) is $T^2$, coordinatized by $\phi$ and the Euclidean
time $\tau \equiv -i t$, with appropriate identifications.  For the
\ads\ solution, the identifications are
\be
(\tau,\phi) \equiv (\tau +\beta,\phi + 2\pi)
\ee 
where $\beta$ denotes inverse temperature and is arbitrary. For the
BTZ solution (with mass $M$ and Euclidean angular momentum $
J_E=iJ$) the identifications are (dictated by smoothness of metric)
\be  
(\tau, \phi) \equiv (\tau + \beta_0, \phi
+\Phi)
\ee 
where $\beta_0 $ and $\phi$ are given by
\bea
\beta_0  &=&\frac{2\pi r_+ l^2}{r_+^2 - r_-^2}  \\ \nonumber
\Phi &=& \frac{2\pi |r_-| l^2}{r_+^2 -r_-^2}  \\ \nonumber
r_+ &=& \left[ \frac{l^2 M}{2} 
\left(1 +\sqrt{1 + \frac{J_E^2}{M^2
l^2}} \right) \right] ^{1/2} \\ \nonumber
r_- &=& -i \left[ \frac{l^2 M}{2} \left( 
\sqrt {1 +  \frac{J_E^2}{M^2
l^2}}-1 \right) \right]^{1/2} \\ \nonumber
\eea
In anticipation we note that implication of the identifications is
that the boundary geometry is that of a torus with the modular
parameter proportional to $\beta_0 + i \Phi $. Therefore the partition
function of the dual CFT need to be evaluated on the torus with the
modular parameter proportional to $\beta_0 + i \Phi$.  The boundary
condition on the fields is that the bosonic fields must have the same
values at identified points whereas the fermion fields must have the
same value upto a sign (periodic (P) or antiperiodic (A)). Since we
are talking about identifications on a two-torus, the fermion boundary
conditions can be \be \alpha = (P, P), (P, A), (A, P), (A, A) \ee
where the first entry denotes boundary condition in the $\phi$
direction and the second entry denotes boundary condition along the
$\tau$ direction.

For the conical spaces, the identification is the same as for \ads,
except that the specification of the functional integral includes the
Wilson line (equation \eq{wilsonline}) as a part of the boundary
condition.

\sgap

\ni\underbar{Saddle points}

\sgap

We will evaluate \eq{free} by finding saddle points of the action
subject to specific boundary conditions. These are Euclidean versions
of the classical solutions that we have described in the previous
section.  By virtue of the equation of motion $R=-6/l^2$, the
Euclidean action $S$ of a classical spacetime $X$ is simply its volume
times a constant. To be precise,
\bea
\label{euclideanaction}
S(X) &=& \frac{1}{4 \pi l^2 G^{(3)}_N} \hbox{Vol}(X)
\nn\\
\hbox{Vol}(X) &=& \int_0^\beta d\tau
\int_{r_0}^R dr \int_0^{2\pi} d\phi \sqrt{g}
\eea
The ranges of $\phi, \tau$ follow from the identifications mentioned
above. The lower limit $r_0$ of the $r$-integral is identically zero
for \ads\ and the conical spaces, whereas for BTZ it denotes the
location of the horizon (the Euclidean section is defined only upto
the horizon). The upper limit $R$ is kept as an infrared regulator to
make the volume finite. We will in practice only be interested in free
energies relative to \ads\ and the $R$-dependent divergent term will
disappear from that calculation.

\sgap

\ni\underbar{Free energy of BTZ:}

\sgap

We will now compute the free energy of BTZ relative to \ads. The \ads\
solution can be at any temperature $\beta $ while the temperature of
the black hole is fixed to be $\beta_0$. To compare with the $AdS_3$
background one must adjust $\beta$ so that the geometries of the two
manifolds match at the hypersurface of radius $R$ (in other words, we
must use the {\em same} infrared regulator on all saddle points of the
functional integral). This gives the following relation
\be
\label{betabeta0}
\beta_0 = \beta\sqrt{\frac{1+l^2/R^2}{1-M^2l^2/R^2}}
\ee
\bea
\label{freebtz}
S(BTZ)-S(AdS_3) 
&=& \frac{1}{4 \pi l^2 G^{(3)}_N}\left[ \int_{BTZ} d^3 x
\sqrt{g} - \int_{AdS_3} d^3 x \sqrt{g}\right]  \\ \nonumber
&=& \frac{1}{4 \pi l^2 G^{(3)}_N}\left[\pi \beta_0 (R^2 - r_+^2) -
\pi\beta R^2\right]   
\eea
Substituting the value of $\beta_0$ in terms of $\beta$ and taking
the limit $R\rightarrow\infty$ we obtain
\be
S(BTZ) -S(AdS_3) = \frac{1}{4 \pi l^2 G^{(3)}_N}\left[\pi\beta_0 l^2 -
\pi^2 r_+ l^2\right] 
\ee
For convenience let us define the left and right temperatures 
\cite{MalStr} as
\bea
\beta_+ =\frac{2\pi l^2}{r_+ + r_- } \\   \nonumber
\beta_- = \bar{\beta_+} = \frac{2\pi l^2}{r_+ - r_- }
\eea
Using these variables the difference in the action becomes
\be
\label{freebtzvalue}
S(BTZ) -S(AdS_3) = \frac{1}{4 \pi l^2 G^{(3)}_N}
\left[\frac{\pi}{2}(\beta_+ + \beta_- ) l^2  
- \pi^3 l^4 \left(
\frac{1}{\beta_+ }+ \frac{1}{\beta_- }
\right) \right] 
\ee
It is easily seen that the BTZ black hole dominates when $(1/\beta_+ +
1/\beta_- )$ is much larger than $ (\beta_+ + \beta_- )$ and vice
versa. Thus at high temperatures we can ignore the first term in the
above equation.
\be
S(BTZ) -S(AdS_3) = -  \frac{\pi^2 l^2}{4 G^{(3)}_N} 
\left(\frac{1}{\beta_+ }
+ \frac{1}{\beta_- }
\right)
\ee
Using  equation \eq{actionparameters}, the high temperature 
partition function of BTZ is, therefore, given by
\be
\label{freebtzhigh}
-\ln Z = \pi^2 Q_1 Q_5 l 
\left(\frac{1}{\beta_+} + \frac{1}{\beta_-}\right)
\ee

\sgap

\ni\underbar{Free energy of conical spaces}

\sgap

We denote the conical spaces by $X_\g$. The volume of $X_\g$ is given
by
\be
{\rm Vol}(X_\g) = \int_0^R r dr \int_0^{\beta_\g} d\tau 
\int_0^{2 \pi} d\phi  = \pi \beta_\g R^2
\ee
Once again, $\beta_\g$ is determined in terms of $\beta$ by the
requirement that the hypersurface $r=R$ that acts as an infrared
regulator has the same 2-geometry as the corresponding surface in
\ads. This gives:
\be
\beta_g \sqrt{R^2/l^2 + \g} = \beta\sqrt{R^2/l^2 + 1}
\ee
Using this, it is easy to find
\be
\label{volconic}
{\rm Vol}(X_\g) - {\rm Vol}(AdS_3) = \pi \beta \frac{l^2}{2}
(1-\g)
\ee 
This leads to, by \eq{euclideanaction}, the following expression
for the Euclidean action
\be
\label{freeconic}
S(X_\g) - S(AdS_3) = \frac{\beta}{8 G^{(3)}_N}(1 - \gamma)
\ee
In the next section we will compare with the free energy computed
from the boundary CFT. For that comparison it will turn out to be more
appropriate to consider as reference spacetime the BTZ black hole with
$J=M=0$ (which is simply the space $X_0$, also 
denoted BTZ$_0$):
\be
\label{freeconicnew}
S(X_\g) - S(\hbox{BTZ}_0) 
= -\ln Z(X_\g) - \left(-\ln Z(X_0)
\right)  = -\frac{\beta}{8 G^{(3)}_N}\gamma
\ee
  
\section{The partition function from the  CFT}

The aim of this section is to calculate the partition function of the
$(4,4)$ CFT on the orbifold $T^{4Q_1 Q_5}/S(Q_1 Q_5)$. The partition
function will depend of the boundary conditions of the fermions of the
CFT. Different bulk geometries will induce different boundary
conditions for the fermions of the CFT.  We will first calculate the
partition function when the bulk geometry is that of the BTZ black
hole.

\sgap

\ni\underbar{CFT partition function corresponding to BTZ}

\sgap

The fermions of the CFT are periodic along the angular coordinate of
the cylinder if the bulk geometry is that of the BTZ black hole. This
can be seen by observing that the zero mass BTZ black hole admits
killing vectors which are periodic along the angular coordinate
\cite{CouHen}. Therefore the zero mass BTZ black hole correspond to
the Ramond sector of the CFT. The general case of the BTZ black hole
with mass and angular momentum correspond to excited states of the CFT
over the Ramond vacuum with
\bea
L_{0} + \bar{L}_0 &=& M l \\  \nonumber
L_0 - \bar{L}_0 &=& J_E
\eea
where $M$ and $J_E$ 
are the mass and the (Euclidean) angular momentum of the BTZ black
hole. Therefore the partition function of the BTZ black hole should
correspond to
\be
Z= \mbox{Tr}_R (e^{2\pi i \tau L_0} e^{2\pi i
\bar{\tau} \bar{L}_0} )
\ee
The Hilbert space of the CFT on the orbifold $T^{4Q_1 Q_5}/S(Q_1 Q_5)$
can be decomposed into twisted sectors labeled by the conjugacy
classes of the permutation group $S(Q_1 Q_5)$.  The conjugacy classes
of the permutation group consists of cyclic groups of various
lengths. The various conjugacy classes and the multiplicity in which
they occur in $S(Q_1 Q_5)$ can be found from the solutions of the
equation
\be
\sum_{n=0}^{Q_1Q_5} n N_n = Q_1 Q_5
\ee
where $n$ is the length of the cycle and $N_n$ is the multiplicity
of the cycle. The Hilbert space is given by
\be
{\cal H} = \bigoplus_{\sum n N_n = Q_1 Q_5} \bigotimes_{n>0}
 S^{N_n} {\cal H}_{(n)}^{P_n}
\ee
$S^N {\cal H}$ denotes the symmetrized product of the Hilbert space
${\cal H}$, $N$ times. By the symbol ${\cal H}_{(n)}^{P_n}$ we mean
the Hilbert space of the twisted sector with a cycle of length $n$ in
which only states which are invariant under the  projection
operator 
\be
P_n = \frac{1}{n} \sum _{k=1}^{n} e^{2\pi i k (L_0 - \bar{L}_0)}
\ee
are retained.
The values of $L_0$ or $\bar{L}_0$ in the twisted sector of length $n$
is of the form $p/n$ where $p$ is positive integer. This projection
forces the value of $L_0 -\bar{L}_0$ to be an integer on the twisted
sector. It arises because the black hole can exchange only integer
valued Kaluza-Klein momentum with the bulk \cite{DavManWad}.

The dominant contribution to the partition function arises from the
maximally twisted sector. That is, from the longest single cycle of
length $Q_1 Q_5$. It is given by
\be
Z= \sum_{m,n} d(Q_1 Q_5 n +m) d(m) e^{2 \pi i n \tau} 
e^{2\pi i m \tau /Q_1 Q_5} e^{-2\pi i m \bar{\tau}/Q_1 Q_5}
\ee
Where $d$'s are the coefficients defined by the expansion
\be
Z_{T^4} = \left[ \frac{\Theta_2 (0|\tau }{\eta ^3 (\tau )}\right] ^2
     = \sum_{n\geq 0}  d(n) e^{2 \pi i \tau n}
\ee
In the above equation $Z_{T^4}$ is the partition function of the
holomorphic sector of the CFT on $T^4$. We will first evaluate the sum
\be
P(m, \tau) = \sum _{n=0} ^{\infty} d(Q_1 Q_5 n + m) e^{2 \pi i n \tau}
\ee
For large values of $Q_1 Q_5 $ we can use the asymptotic form of
$d(Q_1 Q_5 n +m)$
\be
d(Q_1 Q_5 n + m) \sim \exp\left(2 \pi \sqrt{Q_1 Q_5 n +m}\right)
\ee
Substituting the above value of $d(Q_1 Q_5 n +m )$ in $P(m, \tau )$ we
obtain a sum which has an integral representation as shown below.
\bea
P(m, \tau) &=& \sum_{n=1}^{\infty} e^{2 \pi \sqrt{Q_1 Q_5 n + m} + 
2\pi i n \tau } + d(m) \\     \nonumber
&=& {\cal P}\frac{i}{2} \int_{-\infty}^{\infty} dw \coth \pi\omega 
e^{2\pi \sqrt{i Q_1 Q_5 \omega + m } -2 \pi \omega \tau}  + 
d (m) -\frac{e^{2\pi \sqrt m}}{2}
\eea
where ${\cal P}$ denotes ``principal value'' of the integral.
 
We are interested in the high temperature limit of the partition
function. The leading contribution to the integral in the limit
$\tau\rightarrow 0$ is
\be
P(m, \tau) \sim \sqrt{i\pi Q_1 Q_5/ \tau}
e^{i \pi Q_1 Q_5/2\tau -i 2 \pi m \tau/Q_1 Q_5}
\ee
Substituting the above value of $P(m, \tau)$ the partition function
becomes
\be
Z= \sqrt{i\pi Q_1 Q_5/ \tau}  
\sum _{m=0}^{\infty} d(m) e^{-2
\pi i m \bar{\tau}/Q_1 Q_5} 
\sim \exp\left(i \pi Q_1 Q_5 (1/2\tau -1/2\bar{\tau}) \right)
\ee
Thus the free energy at high temperatures is given by
\be
-\ln Z = \frac{-i \pi Q_1 Q_5}{2} 
\left(\frac{1}{\tau} -\frac{1}{\bar{\tau}}
\right)
\ee
This exactly agrees with \eq{freebtzhigh}\ with the 
identification $\tau = i\beta_+/(2 \pi l)$. 

\sgap

\ni\underbar{CFT partition function corresponding to conical
spaces}

\sgap

We will focus on the low temperature (large $\beta$) 
behaviour. At low temperature, using $\tau= i\beta/(2 \pi l)$
($\beta$ real in this case), the partition function is
given by
\be
Z \equiv {\rm Tr} \exp[2 \pi i \tau L_0 - 2 \pi i \bar{\tau}\bar L_0)]
= \exp[-\beta(E_0 + \bar E_0)] (1 + O(\exp[-\beta \Delta]))
\ee
Where $E_0, \bar E_0$ represent the ground state values
of $L_0, \bar L_0$, and $\Delta$ represents the
excitation energy of the first excited 
level.  According to our proposal, the boundary
CFT for conical spaces corresponds to the 
Hilbert space of spectral flow satisfying the relation
\eq{etagamma}. By using \eq{g_energy} and
\eq{landl}, we find that
\be
-\ln Z = \frac{\beta(E_0 + \bar E_0)}{l} = \beta M
= \frac{\beta}{8 G_N^{(3)}} (-\g)
\ee 
which agrees exactly with \eq{freeconicnew} (note that
for the zero-mass BTZ solution the above expression
vanishes).

\section{Concluding remarks}

We have argued that point mass geometries in three dimensional quantum
gravity with negative cosmological constant are part of the phenomenon
of AdS/CFT correspondence. The boundary CFT corresponds to spectral
flow with the spectral flow parameter identified appropriately
with the point mass, or equivalently (\eq{etagamma})  with the defect
angle of the conical singularity.

The above identification also leads to a correspondence between vertex
operators of the boundary CFT with states in the bulk that corresponds
to the point mass, thus leading to a possible description of
scattering of such point masses in terms of CFT correlations. In this
context it would be very useful to connect with the work of Steif
\cite{Steif} where multi-point-mass classical solutions were
discussed.  In particular, the spatial geometry of such solutions were
obtained by quotienting two dimensional hyperbolic space by
appropriate sub-groups of $SL(2,R)$. For example, quotienting by
sub-groups that generate elliptic isometries gives rise to point
particles, while those with hyperbolic isometries lead to black holes.

The interesting thing about these multi-body solutions is that they
are generally not static, and therefore can be used to study black
hole formation by collision of point particles, or more complicated
processes like collisions of black holes. In the light of the AdS/CFT
correspondence, it would be interesting to know how such processes
manifest themselves in the CFT picture. More precisely, what effect
does quotienting the spatial slice have on the CFT? We have found a
hint to this answer in this paper, for the simple case of a single
point particle: it corresponds to a certain vertex operator in the
SCFT (representation of a point mass in the bulk as vertex operator
in a boundary Liouville theory has been discussed recently in
\cite{Nak}). The understanding of more general quotients holds the
promise of providing insights into some very interesting processes in
(2+1) dimensional gravity.

Another interesting point to emerge in this paper is that the
partition function of the dual CFT agrees exactly with that of the
supergravity at high temperatures. This is remarkable as the partition
function of the CFT was computed in weak coupling and the supergravity
result is expected to be a prediction for the CFT result at strong
coupling.  Note that this agreement works for arbitrary angular
momentum, in particular for $J=0$ which is far from extremality and
hence is far from the BPS limit.

\lgap

Acknowledgement: We would like to thank P. Townsend for useful
discussions.  S.W. would like to tahnk the ASICTP for their warm
hospitality.

\end{document}